\documentclass[%
 reprint,
 amsmath,amssymb,
 aps,
]{revtex4-2}

\usepackage{graphicx}
\usepackage{dcolumn}
\usepackage{bm}
\usepackage{braket}

\begin{document}

\preprint{APS/123-QED}

\title{Saturable Quantum Speed Limits for Imaginary-Time Evolution
}

\author{Kohei Kobayashi}
 \affiliation{Global Research Center for  Quantum Information Science, National Institute of Informatics, 2-1-2 Hitotsubashi,  Chiyoda-ku, Tokyo 101-8340, Japan.}

\date{\today}

\begin{abstract}
We derive a Geometric quantum speed limit (QSL) for imaginary-time evolution, where the dynamics is governed by a non-unitary Schrödinger equation. 
By introducing a cost function based on the angular distance between the normalized evolving state and the initial state, we obtain a lower bound on the evolution time expressed as the ratio between this angle and the time-averaged energy dispersion. 
Our bound is analytical, general, and applicable to arbitrary time-independent Hamiltonians.
We analytically evaluate this bound for two physically motivated cases.
First, we apply it to a two-level system and derive an expression for the minimal time.
Second, we analyze the imaginary-time version of Grover’s quantum search problem and rigorously reproduce the well-known logarithmic scaling $T=\mathcal{ O}(\log N)$ within our QSL framework. 

\end{abstract}

\maketitle


\section{Introduction}

Quantum speed limits (QSLs) set fundamental lower bounds on the time required for a quantum system to evolve from one state to another distinguishable state. 
These bounds play a central role in quantum control 
\cite{Deffner, Caneva, Hegerfeldt, Brouzos}, 
quantum metrology \cite{Taddei, Gessner, Jones, Zwierz},
and quantum computation \cite{Lloyd, Keisuke}, providing both theoretical and practical constraints on how fast a given quantum process can be implemented.
Traditionally, most studies on QSLs have focused on unitary dynamics generated by Hermitian Hamiltonians, leading to well-known results such as the Mandelstam–Tamm and Margolus–Levitin inequalities \cite{MT, ML}. 
In recent years, significant effort has been devoted to generalizing QSLs to non-unitary settings, including open quantum systems \cite{Deffner, Taddei, Campo, Zhang, Pires, Deffner2, Kohei} and measurement-induced dynamics \cite{Pintos, Hou}.

Beyond these conventional frameworks, another class of time evolution has gained increasing attention; 
imaginary-time evolution (ITE) \cite{ITE1, ITE2, ITE3, ITE4, ITE5, ITE6, ITE7, ITE8, ITE9, Okada}. 
In this formalism, time $t$ is replaced with an imaginary parameter $it$, leading to non-unitary dynamics governed by non-Hermitian operators.
Implementing ITE on quantum processors has been explored along several complementary directions.
Variational formulations \cite{ITE1} and the quantum ITE framework \cite{ITE2} approximate non-unitary dynamics by short-depth unitary updates.
Those techniques enable hardware-efficient realizations that have already been applied to chemistry and nuclear structure problems \cite{ITE3}.
A second line employs probabilistic or measurement-induced constructions to enact non-unitary maps;
ancilla-assisted forward and backward real-time schemes \cite{ITE4}, 
non-unitary circuit primitives \cite{ITE5}, and acceleration techniques for probabilistic ITE \cite{ITE6, ITE7}.
In parallel, unitary block-encoding and quantum signal processing provide a general method for simulating non-unitary dynamics within a unitary framework \cite{ITE8}.
Resource-aware strategies such as fragmented ITE further tailor implementations to early-stage hardware \cite{ITE9}.

Despite this surge of interest in ITE, the fundamental runtime limits of such processes have remained largely unexplored.
Existing QSL formulations for unitary or general open-system dynamics do not directly apply to ITE; the non-unitary, norm-changing nature of ITE invalidates the Fubini–Study metric arguments used in the unitary case, and open-system QSLs often yield bounds that are far from tight in the exponentially relaxing regime of ITE.
Yet, knowing the exact or tight runtime bound for ITE is crucial for:
(i) assessing the ultimate efficiency of ground-state preparation and imaginary-time quantum algorithms,
(ii) guiding the design of optimal scheduling functions that approach these limits, and
(iii) providing hardware-independent performance guarantees.

In this work, we address the above problems by developing a geometric QSL for ITE.
By introducing a cost function based on the angular distance between the normalized evolving state and the initial state, 
we derive an explicit lower bound expressed in terms of the instantaneous energy variance.
We further identify Hamiltonian schedules that exactly saturate this bound, thereby showing that it is not only rigorous but also achievable.

To demonstrate the validity and usefulness of our result, 
we apply it to two analytically solvable examples.
First, we consider a two-level system undergoing linear imaginary-time annealing and derive a closed-form expression for the minimal annealing time in terms of the energy gap and angular velocity.
Second, we analyze the Grover search algorithm in ITE and rigorously reproduce the well-known logarithmic scaling $T = \mathcal{O}(\log N)$ \cite{Okada} for the optimal search time using our QSL expression.

\section{Geometric Quantum Speed Limit for ITE}

\subsection{Setup and derivation of QSL}

We consider a quantum state $|\psi(t)\rangle$ evolving under imaginary-time Schrödinger dynamics:
\begin{equation}
\frac{d}{dt} |\psi(t)\rangle = -H(t) |\psi(t)\rangle, 
\label{ite}
\end{equation}
where $H(t)$ is a time-dependent Hamiltonian and 
the initial state is pure $ |\psi(0)\rangle =|\psi_0\rangle$ .
This evolution is non-unitary and causes the norm of the state to decrease over time, driving the system toward the ground state of $H(t)$.

To quantify the speed of evolution in this non-unitary setting, 
we define the normalized evolving state
\begin{equation}
|\phi(t)\rangle := \frac{ |\psi(t)\rangle }{ \|\psi(t)\|},
\end{equation}
which obeys the following equation:
\begin{equation}
\frac{d}{dt} |\phi(t)\rangle = (-H+\langle H \rangle_t )|\phi(t)\rangle,
\end{equation}
where $\langle H \rangle_t =\langle \phi(t)|H|\phi(t)\rangle$ and $|\phi(0)\rangle = |\psi_0\rangle$.

Also we introduce the following cost function 
based on angular distance:
\begin{equation}
\Theta(t) := \arccos\left\{\left| \langle \psi_0 |\phi(t) \rangle \right| \right\},
\label{cost}
\end{equation}
where $0\le \Theta(t) \le \pi/2$.
This takes $\pi/2$ when $|\phi(t)\rangle$ is orthogonal to $|\psi_0\rangle$, and $\Theta(t)=0$ is achived only when $|\phi(t)\rangle=|\psi_0\rangle$.
The cost measures the angle between the normalized current state and the initial state.
Here we derive a lower bound of the time $T$, needed for the angle to evolve from $\Theta(0)=0$ to $\Theta(T)\in(0, \pi/2]$.

We adopt a geometric approach based on the Fubini–Study angle between pure quantum states.
The actual trajectory traversed by the system during the evolution has a length
\begin{equation}
\mathcal{L} := \int_0^T \left\| \frac{d}{dt} |\phi(t)\rangle \right\| dt,
\end{equation}
where $\||x\rangle\|=\sqrt{\langle x|x\rangle}$ is the Euclidean norm.
From the geometry of quantum state space, it is known that the Fubini–Study angle is no greater than the length of any path connecting the two states \cite{Geometry}:
\begin{equation}
\Theta(T) \le \mathcal{L}.
\end{equation}
Therefore, we obtain the inequality
\begin{equation}
T \ge \frac{ \Theta(T) }{ \overline{v} }, 
\end{equation}
where we defined the velocity term 
\begin{equation}
 \overline{v} := \frac{1}{T} \int_0^T \left\| \frac{d}{dt} |\phi(t)\rangle \right\| dt.
\end{equation}

In this case, the velocity term becomes
\begin{align}
\left\| \frac{d}{dt} |\phi(t)\rangle \right\| 
&=\|(-H+\langle H\rangle_t)|\phi(t)\rangle\|     \nonumber \\
&= \sqrt{ \langle H^2 \rangle_t - \langle H \rangle_t^2 } \nonumber \\
&=\Delta H(t),
\end{align}
which is the standard deviation of the Hamiltonian.

Substituting this into the average speed $\overline{v}$, 
we obtain the following lower bound:
\begin{equation}
T \ge \frac{ \Theta(T) }{ \overline{ \Delta H } }.
\end{equation}
This bound is our QSL, giving a geometrically intuitive and operationally meaningful lower bound on the runtime of imaginary-time evolution, purely based on the distinguishability between the initial and final states and the average energy fluctuation during the process.

This bound exhibits several strengths as a geometric 
speed limit in the non-unitary setting. 

First, it naturally extends traditional unitary QSLs to ITE by focusing on the normalized trajectory of the state. 
The cost function $\Theta(t)$ provides a clear geometric measure of distinguishability between the initial and time-evolved states, 
and remains meaningful even when the norm of the state decays.

Second, the bound involves the time-averaged 
energy dispersion $\overline{\Delta H}$, 
which directly connects the dynamical speed of evolution to the variance of the driving Hamiltonian. 
This offers an intuitive interpretation in terms of energetic resource consumption, 
which is particularly relevant in quantum optimization contexts such as imaginary-time annealing or variational quantum algorithms.

Moreover, the bound is achievable; 
it becomes an equality when the evolution takes place within a two-dimensional subspace spanned by the initial and orthogonal states, 
and the angle $\Theta(t)$ increases at a rate exactly given by the instantaneous energy dispersion $\Delta H(t)$. 
Such saturating evolutions can be explicitly constructed and serve as theoretical benchmarks for optimal imaginary-time control protocols.
We discuss a general condition for saturating the bound in the next subsection.

On the other hand, the current formulation is limited to pure initial states, 
and generalization to mixed-state dynamics or open-system settings remains an open challenge. 
Furthermore, while the cost function $\Theta(t)$ represents the angular change between states, 
it does not distinguish phase factor in general. 
Finally, evaluating $\Delta H(t)$ may be costly in practice, especially for large-scale time-dependent systems, 
and care must be taken in numerical implementations.

\subsection{General saturation condition}

Let $P_\perp(t):=I-|\phi(t)\rangle\langle\phi(t)|$
be the projector onto the tangent space at $|\phi(t)\rangle$.
The instantaneous speed in the Fubini-Study metric is given by
\begin{equation}
\|P_\perp(t)\,(-H+\langle H\rangle_t)|\phi(t)\rangle\|=\Delta H(t).
\end{equation}

The inequality $\Theta(T) \le \int_0^T \Delta H(t)dt$
is saturated if and only if the following condition 
holds for almost all $t$:
\begin{equation}
P_\perp(t)\,(-H+\langle H\rangle_t)|\phi(t)\rangle
\ \parallel\ P_\perp(t)\,|\psi_0\rangle,
\end{equation}
with a negative real proportionality constant in the gauge $\langle\psi_0|\phi(t)\rangle\ge 0$.
Equivalently, there exists $\lambda(t)\ge 0$ such that
\begin{equation}
P_\perp(t)\,(-H+\langle H\rangle_t)|\phi(t)\rangle
= -\,\lambda(t)\ \frac{P_\perp(t)|\psi_0\rangle}{\|P_\perp(t)|\psi_0\rangle\|}.
\end{equation}
In this case $\frac{d\Theta(t)}{dt}=\Delta H(t)$ pointwise, 
and integration gives
$\Theta(T)=\int_0^T\Delta H(t)\,dt$.

The proof follows from differentiating $\Theta(t)$ with the phase choice $\langle\psi_0|\phi(t)\rangle\ge0$:
\begin{align}
\frac{d\Theta(t)}{dt}
&= -\frac{ \Re\{\langle\psi_0| \dot \phi(t)\rangle\} }{\sin\Theta(t)}\notag \\
&= -\frac{\Re\,\langle\psi_0|(-H+\langle H\rangle_t)|\phi(t)\rangle}{\sin\Theta(t)}.
\end{align}
Here, because $P_\perp|\dot{\phi}(t)\rangle=(-H+\langle H\rangle_t)|\phi(t)\rangle$,
\begin{equation}
\frac{d\Theta(t)}{dt}
= -\frac{\Re\langle\psi_0|P_\perp(-H+\langle H\rangle_t)|\phi(t)\rangle}{\sin\Theta(t)}.
\end{equation}
By the Cauchy–Schwarz inequality,
\begin{align}
\left|\frac{d\Theta(t)}{dt}\right|\ &\le\ 
\frac{\|P_\perp|\psi_0\rangle\|}{\sin\Theta(t)}
\|P_\perp(-H+\langle H\rangle_t)|\phi\rangle\| \notag \\
&= \Delta H(t),
\end{align}
because $\|P_\perp|\psi_0\rangle\|=\sqrt{\langle \psi_0|P_\perp|\psi_0\rangle}=\sin\Theta(t)$.

Equality in Cauchy–Schwarz occurs if and only if the two tangent vectors 
$P_\perp(t)(-H+\langle H\rangle_t)|\phi(t)\rangle$ 
and $P_\perp(t)|\psi_0\rangle$
are linearly dependent, i.e., parallel or anti-parallel in the tangent space.  
The gauge $\langle\psi_0|\phi(t)\rangle\ge 0$ 
fixes the sign of $\Theta(t)$, 
so that approaching the target state corresponds to $\frac{d\Theta(t)}{dt} < 0$.  
This requires the proportionality constant to 
be a negative real number, ensuring that the velocity vector points exactly along the geodesic toward $|\psi_0\rangle$ in projective Hilbert space.  
Geometrically, the saturation condition therefore means 
that the instantaneous motion lies entirely in the direction of the shortest geodesic connecting $|\phi(t)\rangle$ to $|\psi_0\rangle$, 
with no transverse component in the tangent space.

A simple sufficient condition is that the dynamics is confined to a fixed two-dimensional subspace spanned by $\{|\psi_0\rangle,|\phi(t)\rangle\}$ 
and the projected generator $P_\perp(-H+\langle H\rangle_t)|\phi(t)\rangle$ 
is proportional to $P_\perp|\psi_0\rangle$ at all times. 
This is the case, for example, for diagonal two-level Hamiltonians 
$H=E_a|a\rangle\langle a|+E_b|b\rangle\langle b|$ or rank-1 projectors $H(t)=g(t)\,|a\rangle\langle a|$ in a suitable basis. 
Our examples in the next sections are special cases of this condition.

\section{Example}
To illustrate the saturation of the obtained QSL,
we apply it to two analytically solvable scenarios; 
a two-level system and the Grover search problem.

\subsection{Two-level system}
We consider a two-level system where the excited and ground states are explicitly defined as $|0\rangle=(1,0)^\top$ and $|1\rangle=(0, 1)^\top$, respectively.

We define the Hamiltonian as
\begin{equation}
H = E\, |0\rangle \langle 0|, 
\end{equation}
where $E>0$. 
$H$ assigns energy \(E>0\) to the excited state and $0$ to the ground state.

Let the initial state be a superposition of excited and ground states:
\begin{equation}
|\psi_0\rangle = \cos\theta|0\rangle + \sin\theta|1\rangle, 
\end{equation}
where $0 < \theta < \pi/2$.
Under imaginary-time Schrödinger dynamics, the state evolves as
\begin{equation}
|\psi(t)\rangle = \cos\theta\, e^{-E t}\, |0\rangle + \sin\theta\, |1\rangle.
\end{equation}

This evolution causes the excited-state component 
to decay exponentially while the component of the ground state remains unchanged. 
As a result, the system relaxes to the ground state \(|1\rangle\) in the long-time limit.

The normalized state is given by
\begin{equation}
|\phi(t)\rangle = \frac{|\psi(t)\rangle}{\|\psi(t)\|} 
= \frac{ \cos\theta\, e^{-E t}\, |0\rangle + \sin\theta\, |1\rangle }
{ \sqrt{ \cos^2\theta\, e^{-2E t} + \sin^2\theta } }.
\end{equation}

The angular distance from the initial state is
\begin{equation}
\Theta(t) = \arccos \left( \frac{ \cos^2\theta\, e^{-E t} + \sin^2\theta }
{ \sqrt{ \cos^2\theta\, e^{-2E t} + \sin^2\theta } } \right).
\end{equation}

Meanwhile, the instantaneous energy dispersion is
\begin{align}
\Delta H(t) &=\langle \phi(t)|H^2|\phi(t)\rangle-\langle \phi(t)|H|\phi(t)\rangle^2\notag \\
&=  \frac{ E\sin\theta \cos\theta\, e^{-E t} }
{ \cos^2\theta\, e^{-2E t} + \sin^2\theta }.
\end{align}

We compute the time integral of the dispersion:
\begin{align}
\int_0^T \Delta H(t)\, dt 
&= E \sin\theta \cos\theta \int_0^T 
\frac{ e^{-E t} }{ \cos^2\theta\, e^{-2E t} + \sin^2\theta } dt \notag \\
&= \theta - \arctan(e^{-E T} \cot\theta).
\end{align}
We give its derivation in Appendix A.

On the other hand, letting \(\phi := \arctan(e^{-E T} \cot\theta)\),
we find
\begin{align}
&\cos[\theta - \arctan(e^{-E T} \cot\theta) ] \notag \\
&=\cos\theta\cos\phi+\sin\theta\sin\phi  \notag \\
&= \frac{\sin\theta \cos\theta + \sin\theta \cos\theta e^{-Et} }{\sqrt{\cos^2\theta e^{-2Et}+\sin^2\theta}}.
\end{align}

In particular, when we set \(\theta = \pi/4\), the expression further simplifies, and we find
\begin{equation}
\Theta(T) = \int_0^T \Delta H(t)\, dt,
\end{equation}
which implies that the QSL~(\ref{qsl}) is exactly saturated:
\begin{equation}
T = \frac{\Theta(T)}{ \overline{\Delta H} }.
\end{equation}

This example demonstrates that our QSL 
can be tight for appropriately chosen initial states and Hamiltonians.
The exact saturation of the geometric QSL in the case of $\theta = \pi/4$ is not coincidental, 
but rather a consequence of the symmetry and geometry of the problem.

First, note that the ITE suppresses the excited-state component exponentially:
\begin{equation}
|\psi(t)\rangle = \frac{1}{\sqrt{2}}\, e^{-E t} |0\rangle + \frac{1}{\sqrt{2}}\, |1\rangle.
\end{equation}
This leads to a gradual shift of the state vector from the equal superposition toward the ground state $|1\rangle$.

The normalized state becomes
\begin{equation}
|\phi(t)\rangle 
= \frac{ |0\rangle + e^{-Et} |1\rangle }
{ \sqrt{ 1 + e^{-2Et} } },
\end{equation}
which traces out a trajectory on the surface of the Bloch sphere.
The QSL inequality $\Theta(T) \le \int^T_0 \Delta H(t)dt$ simply reflects that the geodesic distance on the Bloch sphere is no greater than the actual path length of the evolution.
When $\theta=\pi/4$ the dynamics follows a great-circle geodesic directly toward 
$|1\rangle$, so the two lengths coincide and the bound is saturated.

\subsection{Grover algorithm in ITE}

We consider the Hamiltonian
\begin{equation}
H = E_w\,|w\rangle\langle w|+ 
E_\perp\,|w_\perp\rangle\langle w_\perp|,
\label{eq:Hproj}
\end{equation}
where $|w\rangle$ is the marked state in an unstructured search problem and
$|w_\perp\rangle$ is its orthogonal complement.
The constants $E_w$ and $E_\perp$ are the corresponding eigenvalues.
Here $|w_\perp\rangle\langle w_\perp|$ denotes the projector onto the subspace orthogonal to $|w\rangle$, explicitly given by
\begin{equation}
|w_\perp\rangle\langle w_\perp| = I - |w\rangle\langle w|.
\end{equation}
This ensures that $|w\rangle$ is an eigenstate with eigenvalue $E_w$, and all states orthogonal to $|w\rangle$ are eigenstates with the degenerate eigenvalue $E_\perp$.
The total state evolution is confined to the two-dimensional subspace spanned by $\{|w\rangle, |w_\perp\rangle\}$.
The initial state is taken as
\begin{equation}
|\psi_0\rangle = \frac{1}{\sqrt{N}}\,|w\rangle + \sqrt{\frac{N-1}{N}}\,|w_\perp\rangle.
\end{equation}

We parametrize the normalized state as
\begin{equation}
\label{ga1}
|\phi(t)\rangle = \cos\theta(t)\,|w\rangle + \sin\theta(t)\,|w_\perp\rangle,
\end{equation}
and $\tan\theta(0) =\sqrt{N-1}$.
Under ITE, the unnormalized state evolves as
\begin{equation}
\label{ga2}
|\psi(t)\rangle = \cos\theta(0)\, e^{-E_w t} |w\rangle 
+ \sin\theta(0)\, e^{-E_\perp t} |w_\perp\rangle.
\end{equation}

From Eqs (\ref{ga1}) and (\ref{ga2}), we have
\begin{align}
\cos\theta(t) &= 
\frac{\cos\theta(0)\, e^{-E_w t}}
{\sqrt{\cos^2\theta(0)\, e^{-2E_w t} + \sin^2\theta(0)\, e^{-2E_\perp t}}}, \\
\sin\theta(t) &= 
\frac{\sin\theta(0)\, e^{-E_\perp t}}
{\sqrt{\cos^2\theta(0)\, e^{-2E_w t} + \sin^2\theta(0)\, e^{-2E_\perp t}}}.
\end{align}
Their ratio satisfies
\begin{equation}
\tan\theta(t) = \tan\theta(0)\, e^{-(E_\perp - E_w)t}.
\end{equation}
Differentiating and using 
$\frac{d}{dt}\tan\theta = \sec^2\theta \frac{d\theta}{dt}$ yields
\begin{equation}
\frac{d\theta(t)}{dt} = -g\, \sin\theta(t)\cos\theta(t),
\label{eq:theta-eom}
\end{equation}
where $g := E_\perp - E_w$.
For $E_\perp > E_w$ ($g>0$), $\theta(t)$ decreases monotonically, indicating exponential convergence toward $|w\rangle$ (we give the proof in Appendix B).

The instantaneous energy dispersion is
\begin{equation}
\Delta H(t) = |E_\perp - E_w|\,\sin\theta(t)\cos\theta(t).
\end{equation}
Combining with Eq.~(\ref{eq:theta-eom}) shows
\begin{equation}
\frac{d\theta(t)}{dt} = -\Delta H(t),
\end{equation}
up to the sign of $g$. 
Moreover, the fidelity with the initial state is
\begin{align}
\langle \psi_0|\phi(t)\rangle
&=\cos\theta(0)\cos\theta(t)+\sin\theta(0)\sin\theta(t)  \notag \\
&=\cos[\theta(0)-\theta(t)],
\end{align}
so the angular distance is
\begin{equation}
\Theta(t) = \theta(0) - \theta(t).
\end{equation}
Thus,
\begin{equation}
\frac{d\Theta(t)}{dt} = \Delta H(t),
\end{equation}
implying the QSL is exactly saturated:
\begin{equation}
T = \frac{\Theta(T) }{\overline{\Delta H}}.
\end{equation}
Physically, the evolution follows a geodesic in projective Hilbert space at a constant speed
$\Delta H$.

Finally, defining $r(t) := \tan\theta(t)$ gives
\begin{align}
\frac{dr}{dt}= \sec^2\theta\frac{d\theta}{dt}
= -gr \quad
\Rightarrow\quad \frac{d}{dt}\ln r(t) = -g,
\end{align}
with $r(0)=\sqrt{N-1}$.  
To achieve a target success probability $1-\varepsilon$, we require $r(T) \le \varepsilon$, yielding
\begin{equation}
T \ge \frac{1}{g}\left[ \frac12\ln(N-1) + \ln\frac{1}{\varepsilon} \right]
\simeq \frac{1}{g}\left[ \frac12\ln N + \ln\frac{1}{\varepsilon} \right],
\end{equation}
where the last step holds for $N\gg 1$.  
Thus, the runtime scales as $T=\mathcal{O}(\log N)$, much faster than the $\mathcal{O}(\sqrt{N})$ scaling of unitary Grover search.

The above example shows that by designing $H(t)$ so that the motion in projective Hilbert space is along a geodesic at constant speed, 
the geometric QSL for ITE can be exactly saturated. 
While such non-unitary evolution cannot be realized within the standard unitary query model without additional resources, 
it can be implemented in analog settings such as dissipative quantum computing, engineered cooling, or postselection-based protocols.

\section{Conclusion}
We have derived a geometric QSL for ITE of pure states, 
formulated in terms of the angular distance between the normalized evolving state and the initial state.
In contrast to traditional QSLs for unitary dynamics, 
our bound explicitly incorporates the norm-nonconserving nature of ITE and is expressed through the time-averaged energy dispersion normalized by the instantaneous state norm.

The resulting QSL gives an achievable, analytically tractable lower bound on the minimum time required to reach a target fidelity with respect to the initial state.
Its fully geometric form, independent of variational ansätze or numerical optimization, makes it particularly suitable for analytical estimates of algorithmic complexity under idealized conditions.

We demonstrated its applicability through two examples.
For a two-level system, we derived a closed-form bound on the minimal evolution time to achieve a desired ground-state population, expressed directly in terms of the energy gap and the instantaneous state trajectory.
For the ITE version of Grover’s search algorithm, our QSL reproduced the well-known logarithmic scaling 
$T=\mathcal{ O}(\log N)$, confirming its ability to capture the optimal scaling of quantum optimization schemes based on imaginary-time dynamics.

The geometric nature of our framework opens the way to further developments.
Possible directions include extending the formulation to noisy or open-system ITE, and adapting it to variational implementations.
More broadly, the explicit link between state distinguishability, energy fluctuations, and optimization speed in non-unitary quantum algorithms may provide new insights into the fundamental limits of quantum control and computational complexity in realistic settings.

In addition, we have identified a general saturation condition for our QSL, specifying when the instantaneous evolution direction remains perfectly aligned with the geodesic toward the target state.
This condition not only explains when the bound becomes tight but also offers a constructive guideline for designing Hamiltonians and schedules that achieve the minimum possible runtime.

\begin{acknowledgments}
This work was supported by MEXT Quantum Leap Flagship Program Grant JPMXS0120351339. 
\end{acknowledgments}

\appendix

\section{Analytical Evaluation of the Time-Integral of Dispersion}

In this appendix, we evaluate the following time integral of the energy dispersion appearing in the QSL derivation:
\begin{equation}
I := \int_0^T 
\frac{ e^{-E t} }{ \cos^2\theta\, e^{-2E t} + \sin^2\theta } dt.
\end{equation}
We proceed by a change of variable. Define \( x := e^{-E t} \), so that
\[
t = -\frac{1}{E} \ln x, \qquad dt = -\frac{1}{E x} dx.
\]
The integration limits become: 
\( t = 0 \Rightarrow x = 1 \), and \( t = T \Rightarrow x = e^{-E T} \).
Thus, the integral becomes
\begin{align}
I &= \int_{e^{-E T}}^{1}
\frac{x}{\cos^2\theta \cdot x^2 + \sin^2\theta} \cdot \left(-\frac{1}{E x} dx\right) \notag \\
&= \frac{1}{E} \int_{e^{-E T}}^1 \frac{1}{\cos^2\theta \cdot x^2 + \sin^2\theta} dx.
\end{align}

Next, we factor out constants:
\begin{align}
I &= \frac{1}{E \cos^2\theta }
\int_{e^{-E T}}^1 \frac{1}{ x^2 +\tan^2\theta} dx.
\end{align}
We now apply the standard integral formula:
\[
\int \frac{1}{x^2 +a^2} dx = \frac{1}{a} \arctan(a x) + C.
\]
Setting \( a = \tan\theta \), we have:
\begin{align}
I &= \frac{1}{E \cos^2\theta}
\cdot \frac{1}{\tan\theta}
\left[ \arctan(\cot\theta) - \arctan(e^{-E T} \cot\theta) \right] \notag \\
&= \frac{1}{E \sin\theta\cos\theta}
\left[\theta - \arctan(e^{-E T} \cot\theta) \right].
\end{align}

Thus, the full time-integrated energy dispersion is:
\begin{align}
\int_0^T \Delta H(t)\, dt 
&= E \sin\theta \cos\theta \cdot I \notag \\
&= \theta - \arctan(e^{-E T} \cot\theta),
\end{align}
which is the expression used in the main text 
to establish the QSL saturation.

\section{Proof of Convergence to a target state in Grover algorithm}

We analytically show that ITE under a diagonal Hamiltonian with a unique ground state converges to that ground state after normalization, and relate this to the saturation of the geometric QSL.

Consider the Hamiltonian
\begin{equation}
H = E_w\,|w\rangle\langle w| + E_\perp\,\left(I - |w\rangle\langle w|\right),
\label{eq:H_appendix}
\end{equation}
where $E_w, E_\perp \in \mathbb{R}$. 
The state $|w\rangle$ is the target state (candidate ground state), 
and the orthogonal subspace $I - |w\rangle\langle w|$ has degenerate eigenvalue $E_\perp$.

Let the initial state be decomposed as
\begin{equation}
|\psi_0\rangle = \alpha\,|w\rangle + |\chi\rangle, 
\qquad \langle w|\chi\rangle = 0,
\end{equation}
where $\alpha = \langle w|\psi_0\rangle$ and $|\chi\rangle$ is the orthogonal component.

The imaginary-time dynamics
\begin{equation}
\frac{d}{dt}|\psi(t)\rangle = -H\,|\psi(t)\rangle
\end{equation}
has the exact solution
\begin{equation}
|\psi(t)\rangle
= \alpha\,e^{-E_w t}\,|w\rangle + e^{-E_\perp t}\,|\chi\rangle.
\end{equation}
The (unnormalized) excited-state component decays with rate $E_w$ while the orthogonal component decays with rate $E_\perp$.

The squared norm is
\begin{equation}
\||\psi(t)\rangle\|^2 = |\alpha|^2 e^{-2E_w t} + \||\chi\rangle\|^2 e^{-2E_\perp t}.
\end{equation}
The fidelity with $|w\rangle$ for the normalized state 
$|\phi(t)\rangle$ is
\begin{equation}
F(t) := |\langle w|\phi(t)\rangle|^2
= \frac{|\alpha|^2 e^{-2E_w t}}{|\alpha|^2 e^{-2E_w t} + \||\chi\rangle\|^2 e^{-2E_\perp t}}.
\label{eq:fidelity_appendix}
\end{equation}
If $E_w < E_\perp$ (i.e., $|w\rangle$ is the unique ground state) 
and $\alpha \neq 0$, then setting $\Delta := E_\perp - E_w > 0$, as $t\to\infty$, we find
\begin{align}
1-F(t)  
&=  \frac{\frac{\||\chi\rangle\|^2}{|\alpha|^2}
e^{-2\Delta t}}{1 +\frac{\||\chi\rangle\|^2}{|\alpha|^2} e^{-2\Delta t}} \notag \\
&\le \frac{\||\chi\rangle\|^2}{|\alpha|^2} e^{-2\Delta t} \to 0.
\end{align}
Thus, the convergence
\begin{equation}
|\phi(t)\rangle \to |w\rangle
\end{equation}
is realized.

\end{document}